\documentclass[journal]{IEEEtran}

\usepackage{amsmath}
\usepackage{array}
\usepackage{algorithm}
\usepackage{multirow}
\usepackage{graphicx}
\usepackage{epstopdf}
\usepackage{epsfig}
\usepackage{subfigure}
\usepackage{threeparttable}
\usepackage{tabularx}
\usepackage{relsize}
\usepackage{cite}
\usepackage{tikz}
\usepackage{enumerate}

\usepackage[pdftex]{pict2e}
\usepackage{stfloats}
\usepackage{algpseudocode}
\usepackage{csquotes}

\usepackage{amssymb}

\usepackage{cuted}
\usepackage{lipsum}
\DeclareGraphicsExtensions{.pdf, .png, .jpg, .EPS, .tif}

\begin{document}

\title{Odor Intensity Shift Keying (OISK) and Channel Capacity of Odor-based Molecular Communications in Internet of Everything}
    \author{Aditya Powari and Ozgur B. Akan, \IEEEmembership{Fellow, IEEE} 
\vspace{-20pt}

\thanks{
Aditya Powari is with the Department of Electrical and Electronic Engineering, University of Manchester, Manchester M13 9PL, United Kingdom (email: adityapowari@gmail.com).}
\thanks{
Ozgur B. Akan is with the Internet of Everything Group, Electrical
Engineering Division, Department of Engineering, University of
Cambridge, CB3 0FA Cambridge, U.K., and also with the Center for
neXt-Generation Communications (CXC), Department of Electrical and
Electronics Engineering, Koç University, 34450 Istanbul, Turkey (e-mail:
oba21@cam.ac.uk; akan@ku.edu.tr) }
\thanks{
This work was supported in part by the AXA Research Fund (AXA Chair for Internet of Everything at Koç University).}
}

\maketitle

\begin{abstract}
Molecular communication is a new, active area of research that has created a paradigm shift in the way a communication system is perceived. An artificial molecular communication network is created using biological molecules for encoding, transmitting and decoding the symbols to convey information. In addition to typical biological molecules, we are also exploring other classes of molecules that possess unique distinctive features which can be potentially exploited for establishing reliable communications. Odor molecules are one such class of molecules which possess several distinctive features such as Intensity, Headonic tone which provides a basis to convey the information in an olfactory communication system. In our work, we investigate the ICT (information and communication theory) perspective of the olfactory communications by evaluating the channel capacity of an odor molecular communication (OMC) system with the help of a novel modulation scheme viz. odor intensity shift keying (OISK), where information is being conveyed from the intensity level of an odor. Furthermore, we also analyse the effects of critical parameters like temperature and noise on the achievable channel capacity to provide an insight about the resilience of the proposed OMC system towards any such anomaly faced by it.
\end{abstract}
\vspace{-0.4cm}
\section{Introduction}

\IEEEPARstart{N}{atural} olfactory communication is one of the many intriguing mechanisms that is utilized by several living organisms for a wide range of multi-purpose tasks such as sensing and detecting their surroundings, initiating the process of reproduction via the release of sex pheromones (odor based signals), territory marking and even establishing intra/inter species communication. For instance, flowers of a plant release scents which are composed of odor molecules to attract insects or other mobile animals in order to facilitate spreading of their pollen grains. Thus, developing an understanding of the fundamentals governing natural olfactory communication would enable us to potentially develop artificial olfaction based communication networks \cite{Dilara_unpublish}. Research is underway to materialize transceivers having novel architecture, that would make them suitable for odor-based molecular communication (OMC) networks \cite{Kuscu_survey}. For instance, one of the potential candidate for OMC transmitters is the Hydrogel based OMC transmitters \cite{XGuo_Hydrogel_2023},\cite{Nguyen_2017}, which are very promising due to their capability of holding multiple distinctive odors along with having versatile odor emission. Further research could be carried out to improve the geometry of the Hydrogel based OMC transmitters, which could pave the way for achieving more efficient odor encoding and emissions \cite{Covington_2021},\cite{civas_2023graphene},\cite{civas_2022molecular}. 
\\
On the other hand, for OMC receivers the focus is on digitising the sense of smell, i.e., digitising the perception of odors. With the help of techniques like gas chromatography, mass spectrometry these receivers analyze the chemical composition of an odor to enable its detection. Depending on a receivers architectural dimensions along with the detection technique it incorporates, the OMC receivers can be classified into two categories namely: Macro-scale and Micro/Nano-scale designs \cite{Dilara_unpublish}. Some well-known existing designs for the Macro-scale category includes the magnetic nano-particle detector \cite{Bartunik_2019}, mass spectrometry based detectors \cite{Mcguiness_2019},\cite{Mcguiness_2018},\cite{Mcguiness_2019_access} and gas sensor-based receivers \cite{Farsad_2013},\cite{Koo_2016}. These designs require further modifications in order for them to become apt for use as OMC receivers as currently their potential consideration as OMC receivers are hindered by drawbacks such as bulky equipment, limited selectivity and scalability, saturation issues. For the nano-scale based systems some notable existing OMC receiver designs include the Nanomaterial based FETs \cite{Kuscu_2021},\cite{Kuscu_2015},\cite{Kuscu_2016}, \cite{Kuscu_2016_trans}, Mechanical OMC receiver \cite{Dilara_2022},\cite{Dilara_2021}, which utilizes the odor type and odor concentration for implementing odor detection and odor recognition.  
Another promising potential candidate for OMC receiver is the E-nose \cite{Kuscu_2016}. Such systems are composed of chemical array (e.g., metal-oxide array, nano-sensors \cite{KHAN_2020},\cite{Li_2014}) based sensor components that mimics the sense of smell that is inherent in living organisms. An incoming odor stimulates these sensors leading to generation of signals from these sensors. These signals are then analysed and with the help of advanced techniques like pattern recognition and classification the E-nose decodes the information stored in the odor.\\
However, besides the transceiver designs an information communication theory (ICT) perspective, which is crucial for establishing reliable end-to-end transmission of information via odor molecules based signals has not been explored yet. The ICT perspective involves study of several important aspects that are fundamental to any communication system such as minimum signal-to-noise (SNR) requirements, maximum achievable data rate, i.e., channel capacity, effect of interference from multiple simultaneous transmissions, an appropriate $M$-ary modulation scheme for encoding/decoding symbols to name a few. It is imperative to carry out research and gain a comprehensive understanding of OMC systems from an ICT perspective, as it holds the key to unlock the vast potential of artificial olfactory communications, eventually contributing towards realization of Internet of Everything (IoE) \cite{Akan2023MAGAZINE}. In this paper, we investigate the channel capacity of an OMC system and formulate a closed-form analytical solution for the same, whilst also exploring the prospect of harnessing a unique, distinctive feature of odor molecules viz intensity for encoding the symbols, which ultimately leads to the formation of an odor intensity-based signal over space-time.\\
The remainder of this paper is organized as follows. Section
II describes the proposed OMC system model and elaborates in detail the relevant odor propagation mechanisms considered. Section III explains harnessing of a unique odor feature viz intensity to generate an odor intensity signal via the odor-intensity shift-keying (OISK) modulation technique, whereby symbols are encoded/decoded from the intensity of odor. Section IV illustrates the effect of two critical performance governing parameters on the OMC system. Section V presents a generic closed form analytical expression of the channel capacity of an OMC system. Section VI presents the simulation results of the performance of the proposed OMC system model in terms of its achievable capacity in the presence of adverse effects like temperature and noise. Finally, Section VII provides any concluding remarks, potential application avenues of this work along with the required  future work that is essential for the realisation of OMC systems.   
\vspace{-0.3cm}

\section{System Model}
 The natural olfactory communications relies considerably on the presence of natural winds (unbounded environment), which provides advection to the propagating odor molecules. However, the artificial olfactory communications should be designed such that they are also able to operate in a bounded environment, where the necessary airflow for advection is artificially provided. In this section, we describe the essential components and suitable assumptions on the basis of which we propose a novel OMC system model that is suitable for both a bounded as well as an unbounded environment. In the proposed OMC system, we utilize an odor's concentration i.e., intensity, which is discussed later in Section III to exchange information symbols between the transmitter (Tx) and receiver (Rx). To begin with, we assume that the change in the initial emitted odor concentration as it propagates from Tx to Rx, occurs along the direction of odor propagation only. In other words, the change in concentration occurs along one spacial dimension only. Besides the spacial dimension, the change in concentration over time is also considered here. Hence, the proposed OMC system model takes into consideration the change in odor concentration over space-time.\\
 In order to investigate the channel capacity in OMC systems, we first need to define some preliminary parameters that forms the basis of the capacity analysis of the OMC system. Assume that the Transmitter (Tx) is positioned at the location $x=0$ and the Receiver is stationed at say $x=R$, i.e., Tx-Rx separation is $R$, then the change in odor concentration ($\Phi$) takes over space ($x$) and time ($t$). Hence mathematically we denote the odor concentration as a multivariate function viz. $\Phi(x,t)$. We assume a constant, streamline airflow is present along the entire odor propagation path, which results in the advection of odor molecules. The speed of this airflow is denoted by $v$. Depending on the system environment setting, i.e., bounded or unbounded, this airflow could be either provided artificially or be present naturally respectively. Now, as the airflow is constant throughout the propagation path it is safe to convey that time taken for emitted odor to reach Rx would be $R/v$. At the Tx, each information symbol is emitted within a fixed span of time known as the symbol slot duration. In other words, this is in fact the symbol period as used in classical communications. We denoted it here by $\tau$.  

\subsection{Odor Propagation Mechanism}

 In our OMC system, we take into consideration both the \textit{diffusion} as well as the \textit{advection} mechanisms of molecule propagation. Diffusion is a perennial mechanism that governs the odor molecule propagation due to difference in concentration of odor at the transmitter (Tx) and the receiver (Rx). On the other hand, the odor molecules also experience an advection when there is presence of a constant, streamline airflow between Tx and Rx. As stated earlier, this airflow could either be present naturally due to winds (unbounded environment) or generated artificially (bounded environment). In our system model, we consider that both these propagation mechanisms are working simultaneously on the transmitted odor molecules, and hence, the propagation mechanism for the odor molecules is described by an Advection-Diffusion model. \cite{TracerEquation} describes one such model for molecule propagation by the means of tracer equation, which is expressed as:
 \begin{equation}
     \frac{\partial \Phi}{\partial t} = -U.\nabla \Phi + \nabla . K \nabla \Phi
     \label{tracer}
 \end{equation}
 where $\Phi$ is the molecule concentration, $U$ is the airflow velocity and $K$ is the eddy diffusivity coefficient. In this equation the variation of molecule concentration $\Phi$ over the space and time has been modelled by incorporating the eddy diffusivity coefficient $K$ for turbulent conditions in the air. This partial differential equation (PDE) model can be used for a multi-dimensional environment and it describes the change in concentration of molecules present in the air. It is to be noted that Eddy diffusion is a very specific type of diffusion mechanism that occurs only in the presence of turbulent fluid flows, where it tends to dominate over the ordinary diffusion process. Hence, in order to get (\ref{tracer}) suitable for modelling the odor propagation in an OMC system we need to implement some modifications. Firstly, the eddy diffusivity coefficient $K$ has been replaced with the typical ordinary diffusion coefficient $D$ of an odorant. The core reason for this modification is that such turbulent conditions, in which the eddy diffusion coefficient operates, are extremely unlikely to happen in a bounded environment. For an unbounded scenario as well, the occurrence of such turbulent winds are quite rare, and hence, it is not fit to base the odor propagation model for OMC systems upon the eddy diffusivity $K$.

Secondly, as the OMC system model here assumes a one-dimensional odor propagation between Tx and Rx, it implies that the variation in odor concentration occurs over a single space dimension (along $x$ only) along with time ($t$). As a result the gradient operator used in (\ref{tracer}) is extended only along the $x$ direction. Hence, the modified PDE model governing the advection-diffusion mechanisms for the odor molecules of an OMC system can be expressed as:
\vspace{-0.3cm}
\begin{equation}
    \frac{\partial \Phi}{\partial t} +  v\bigg(\frac{\partial \Phi}{\partial x}\bigg) = D \bigg(\frac{\partial^2 \Phi}{\partial^2 x}\bigg)
    \label{PDE}
\end{equation}

 where, $\Phi(x,t)$ represents the variation of concentration over space and time, $D$ is the diffusion coefficient of the odorant used and $v$ is the speed of the airflow creating the advection phenomenon. It is to be noted that $D$ here is a constant entity over propagation distance ($x$) and so directly comes outside the second order derivative. However, $D$ in general varies with other associated entities such as humidity, which makes it a random entity as well, which we would discuss about in more detail later on. 
 \vspace{-0.5cm}
 \subsection{Solution of the Advection-Diffusion PDE model}

 The determination of a general solution for complex PDEs such as (\ref{PDE}) is a tricky task and on several occasions they might not even exist. Hence, it is beneficial to determine an exact solution for a PDE that accurately resembles the variable of interest. In order to obtain an exact solution for the partial differential equation stated in (\ref{PDE}), we need to first define a set of relevant boundary conditions that govern the proposed OMC system model. These boundary conditions directly govern the closed form analytical solution for the variable of interest $\Phi(x,t)$ and are as follows:
 \begin{enumerate}[(1)]
     \item $\Phi(0,0)=A$ implying that at time $t=0$ the  initial concentration of the odor at Tx is $A$.
     \item $\Phi(R,0)=0$ implying that at time $t=0$ there is zero concentration of odor at the Rx which is placed at $x=R$.
    \item $\Phi(0,\tau) = 0$ emphasizing that after the completion of a symbol slot (of duration $\tau$), the odor concentration at Tx will drop to zero again.
 \end{enumerate}
 With these three boundary conditions in place, we can proceed to find out the exact  solution for (\ref{PDE}). For this, we make use of the \textit{Dsolve} function of the \textit{Wolfram MATHEMATICA} software to obtain an exact solution for the PDE model by incorporating the aforementioned boundary conditions. The analytical expression for the solution was obtained as:
 
 \vspace{-0.7cm}
 \begin{multline}\label{Solution}
     \Phi(x,t) = \frac{1}{R\tau(2D+Rv+\tau v^2)} \bigg(-2ADRt+ 2ADR\tau\\ -AR^2tv + AR^2\tau v -ARt^2v^2 +AR\tau^2v^2  -At^2\tau v^3 \\+ At\tau^2 v^3  +AR^2x -2AD\tau x + 2ARtvx \\ + 2At\tau v^2 x - A\tau^2 v^2 x -ARx^2 -A\tau v x^2 \bigg)
 \end{multline}
 
 This solution would help us to understand and analyse how the concentration of the transmitted odor would vary over the space ($x$) , time ($t$) along with its dependence on key parameters like initial concentration ($A$), airflow speed ($v$) and the symbol time slot duration ($\tau$) and the diffusion coefficient ($D$) of the odorant being used. 

 \section{The OISK Modulation Scheme}
\begin{figure}[t!]
     \centering
     \includegraphics[width=3.4in,height=1.8in]{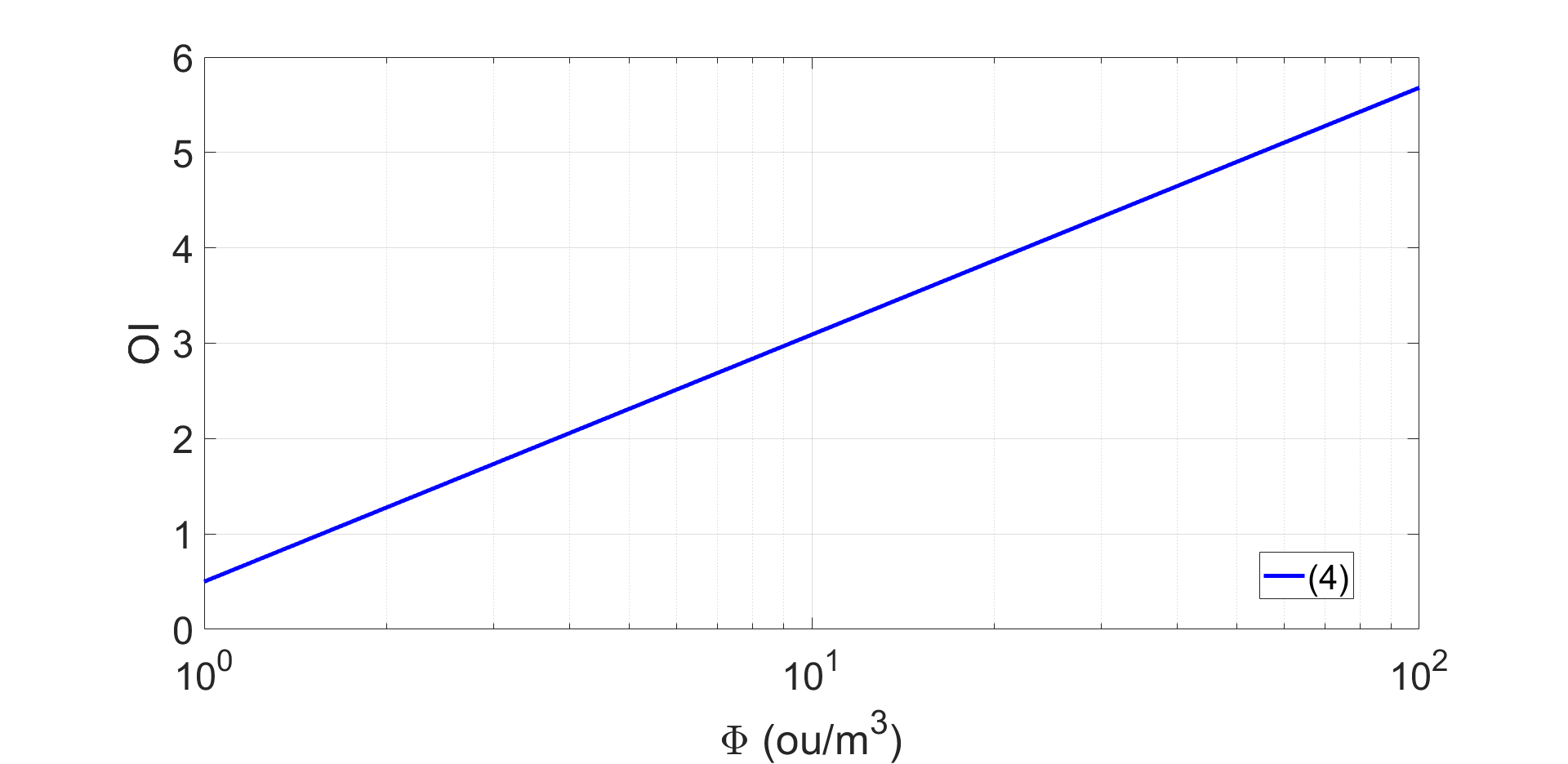}
     \caption{Relationship between odor intensity and odor concentration as per (\ref{Weber_fechner_law}) }
     \label{Weber1}
 \end{figure}

In this paper, we also propose an odor based signal for conveying information in artificial olfactory communication networks. This signal utilizes a distinctive feature of odor viz. odor intensity in order to generate an odor specific message signal. We call this scheme as the Odor Intensity Shift Keying (OISK), where the information contained in an odor symbol is expressed by its intensity. The German standard Olfactometry Determination of Odour Intensity (VDI 3882 Part 1) \cite{GermanVDI} outlines a qualitative scale to measure the odor intensity which is shown in Table \ref{intensity_scale}.
\begin{table}[t!]
    \centering
    \caption{Odor Intensity Scale}
    \label{intensity_scale}
    \scalebox{1.1}{
    \begin{tabular}{|c|c|}
        \hline
       \textbf{\textit{Intensity level}}  & \textbf{\textit{Perception}} \\
       \hline
        0 &  Not perceptible\\
         \hline
        1 &  Very Weak\\
         \hline
        2 & Weak \\
         \hline
        3 & Distinct \\
         \hline
        4 & Strong \\
          \hline
        5 & Very Strong \\
         \hline
        6 & Extremely Strong \\
         \hline
    \end{tabular}}
    \vspace{-0.3cm}
\end{table}
The idea of OISK is to encode each symbol in a unique odor intensity level as highlighted in Table \ref{intensity_scale}. For instance, if the number of symbols in the Tx alphabet set are two, i.e., $M=2$, which implies a binary symbol set consisting of bits \enquote{0} and \enquote{1} only. Now, for each of these binary symbols, we assign them a unique intensity level in order to encode the symbol. The emission of such symbols in a successive manner would lead to the formation of an odor signal over space-time. It is important to also lay emphasis on the relationship between odor intensity and odor concentration, which plays a crucial role for implementing the encoding and decoding of information. The well known Weber-Fechner law establishes a logarithmic relationship between odor concentration ($OC$) having units of $ou/m^3$ where $ou$ stands for odor units and the odor intensity ($OI$). Mathematically this law has been defined in \cite{Weber_Fechner_Coeff} as
\vspace{-0.2cm}
\[
    OI = k \log (\Phi) + d
\]
where $k$ is the Weber-Fechner constant, which is different for different odorants and $d$ is the intensity intercept value which again varies for different odors. Now, \cite{Weber_Fechner_modify} and \cite{GermanVDI} have highlighted that the value of this intercept should be taken as $d=0.5$ , i.e., a fixed value for all odors, reason being that only $50\%$ of the panel members for odor detection were able to perceive odors having a concentration of $\Phi=1$ $ou/m^3$ as weak odor whilst other members could not perceive any odor at all. This leads us to a modified version of the weber-fechner law, which is stated as
\vspace{-0.2cm}
\begin{equation}
    OI = k \log (\Phi) + 0.5
    \label{Weber_fechner_law}
\end{equation}
For our analysis, we have taken the odorant to be benzene, which is an aromatic hydrocarbon based odor. The value of the Weber-Fechner constant $k=2.59$ for benzene odor \cite{Weber_Fechner_Coeff}. We have taken the environment temperature in the surroundings of the OMC system model to be $298$ \textit{K}, at which the pressure-independent diffusion coefficient for benzene is found to be $D=72\pm 3$ \textit{Torr} $cm^2/s$ \cite{Diffusion_coeff}. In other words, the diffusion coefficient is within the range of $D\in[69,75]$ \textit{Torr} $cm^2/s$. In this paper, we also consider the pressure in OMC system environment to be be $1$ \textit{atm} or $760$ \textit{Torr}. Hence, for the OMC system we would have the diffusion coefficient for benzene belonging to the range of $D\in[9.078\times 10^{-6},9.868\times 10^{-6}$] $m^2/s$. This variation in the measured values of $D$ for benzene at $298$ \textit{K} leads to $D$ being the only parameter in OMC system that possess a degree of randomness, and hence, it is logical to treat $D$ as a random variable while evaluating the channel capacity. It is to be noted that this range of $D$ is true only when the OMC system environment has a temperature $T=298$ \textit{K} and a pressure of $1$ \textit{atm}. \\
For benzene, using (\ref{Weber_fechner_law}) with $k=2.59$ and $d=0.5$ the variation of odor intensity with odor concentration has been shown in Fig. \ref{Weber1} using a logarithmic scale for x-axis. From Fig. \ref{Weber1}, it is apparent that an odor intensity level could only be achieved when the corresponding odor concentration is of an appropriate magnitude.

 \subsection{Symbol Encoding and Decoding in OISK}
\begin{figure}[h!]
     \centering
     \includegraphics[width=3.4in,height=1.8in]{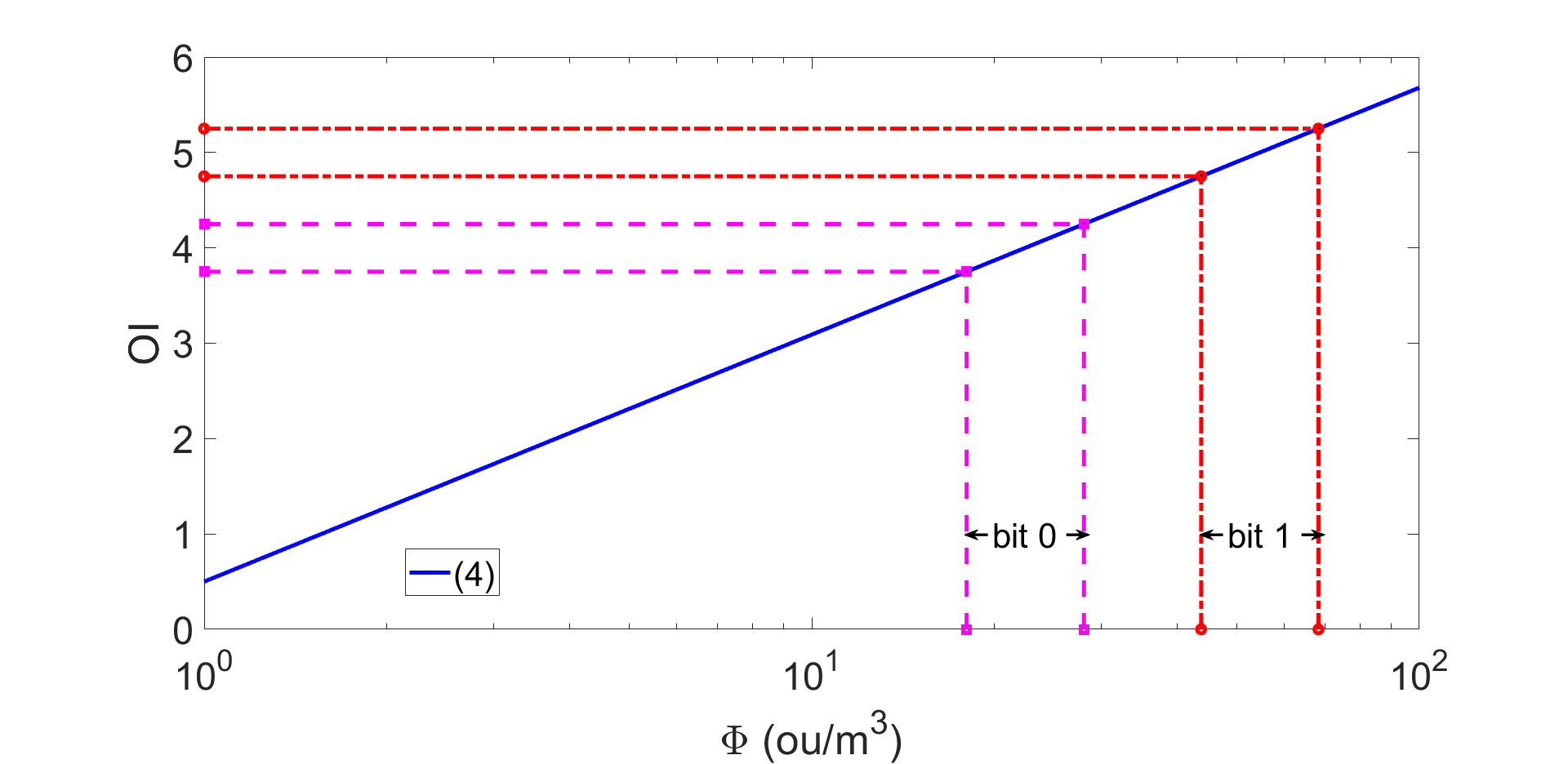}
     \caption{Illustration of corresponding concentration ranges for both intensity levels}
     \label{Weber_bands}
 \end{figure}

 Here, we consider that each odor intensity level is in fact not a unique intensity value that corresponds to a unique concentration but rather its an intensity band that is centered at the value of the intensity level, as illustrated in Fig. \ref{Weber_bands}. The figure depicts that any such intensity band (centered at an intensity level) would correspond to an equivalent band of odor concentration. Let the value of an intensity level be $I$ for $I\in\{0,1,2,3,4,5,6\}$ and hence, the range of the intensity band encompassing this level could be chosen as $I\pm0.25$. In other words, we can say that the width of an intensity band i.e., intensity bandwidth, is equal to $0.5$ here and as a result the corresponding concentration band is also of sufficient width. The two reasons for selecting this intensity bandwidth are as follows:
 \begin{enumerate}[1.]
     \item Firstly, to ensure a suitable gap between adjacent intensity, and, in turn, concentration, bands to avoid errors due to cross symbol detection.
     \item  Secondly, to ensure sufficient width of concentration band, which would enable the OMC system to combat fluctuations in the received odor concentration due to noise induced errors. 
 \end{enumerate}
 The choice of intensity bandwidth used in the OISK modulation is extremely crucial for ensuring systems resilience towards adverse effects like noise and temperature fluctuations in the OMC environment. The importance of the intensity bandwidth would be further highlighted in Section VI when we delve into the analysis of channel capacity.\\ 
 For the OISK scheme with $M=2$ symbols, we have selected intensity levels 4 and 5 to represent the symbol bits \enquote{0} and \enquote{1} respectively. The resulting intensity and concentrations bands arising from this selection are shown in Fig. \ref{Weber_bands}. The range values for the intensity and concentration bands shown in Fig. \ref{Weber_bands} are listed in Table \ref{range_table}.

 \begin{table}[t!]
     \centering
     \caption{Ranges of the Intensity and Concentration bands}
     \label{range_table}
     \scalebox{1.1}{
     \begin{tabular}{|c|c|c|}
        \hline
        \textbf{\textit{Band Type}} & \textbf{\textit{Bit}}  &  \textbf{\textit{Range}}  \\
         \hline
        \multirow{2}{*}{Intensity} & Bit "0" & $[3.75,4.25]$  \\
         \cline{2-3}
         & Bit "1" & $[4.75,5.25]$  \\
         \hline
         \multirow{2}{*}{Concentration} & Bit "0" & $[17.9815,28.0463]$  \\
         \cline{2-3}
         & Bit "1" &  $[43.7448,68.2320]$ \\
         \hline
 \end{tabular}}
 \end{table}
  For encoding, the Tx prepares an odor symbol at its designated odor intensity level by ensuring its concentration is equivalent to the end value of its appropriate concentration range, i.e., $28.0468$ $ou/m^3$ for bit \enquote{0} and $68.2320$ $ou/m^3$ for bit \enquote{1}. To ensure that the correct odor intensity level is decoded, the received concentration of an odor symbol at Rx must be within its specified concentration range as in Table \ref{range_table}.
  Transmitting odor symbol at end value of its concentration range ensures enough margin for the transmitted symbol for it to be received within its appropriate concentration detection range at Rx in the presence adverse effects such as propagation noise and Additive White Gaussian Noise (AWGN), which are discussed next.

 \begin{figure*}[t!]
     \centering
     \includegraphics[width=7.1in,height=2in]{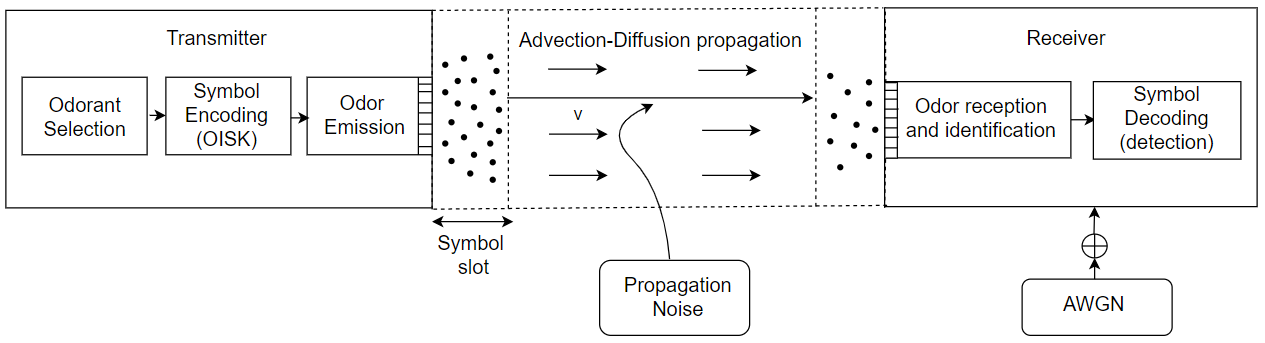}
     \caption{The OMC System Model with inclusion of propagation and AWGN noise }
     \label{OMC_system_diagram}
     \vspace{-0.3cm}
 \end{figure*}
 \subsection{Effects of Propagation Noise and AWGN }
 The advection effect is observed due to the presence of a constant, streamline, directed airflow which helps in guiding the emitted odor molecules from Tx towards the Rx with the airflow speed $v$. However, besides advection, the odor molecule propagation here is also governed by diffusion of the odor molecules. Now, we know that the diffusion refers to the movement of odor molecules from regions of high concentration to regions of low concentration and is driven by the random thermal motion undertaken by the molecules. This random motion of odor molecules leads to loss in the odor concentration over space-time, which results in the received odor concentration being less than the initial transmitted odor concentration. This effect is called the diffusion-induced noise and since it causes deterioration of odor concentration during the propagation, we refer to it as the propagation noise. This noise is characterized by uncertainty in the diffusion coefficient $D$. The presence of a directed airflow can mitigate this effect to a large extent, however it cannot be completely eradicated as diffusion is a perennial mechanism. Due to the propagation noise, fluctuations are observed in received odor concentration at Rx. This concentration tend to follow a Gaussian distribution as a result of the Central Limit Theorem. This theorem is applicable here as we have a large number of diffusing odor molecules, i.e., concentration, that are being used to convey the information via odor symbols.
 Furthermore, at the receiver itself the AWGN would directly affect the magnitude of the received odor concentration which could potentially lead to errors during the symbol decoding procedure. The AWGN at the Rx primarily would arise due to thermal heat induced noise and imperfections in the associated electronic components used in the Rx hardware. Hence, it is important to consider both of these noise types to get a more pragmatic understanding of the OMC system performance in terms of its achievable channel capacity. Both of these adverse noise based effects have been incorporated in our OMC system model for channel capacity evaluation via the use of Monte Carlo simulations.\\
 A schematic diagram of the proposed OMC system model is shown in Fig. \ref{OMC_system_diagram}. The diagram illustrates the necessary steps that are to be carried out by the Tx which eventually leads to the emission of an OISK symbol within the prefixed symbol time slot. The airflow present in the OMC system (either natural wind or artificial airflow) is used to guide the odor molecules towards the receiver. However, due to the propagation noise the received odor concentration is lower that what was transmitted. At the receiver itself, the presence of AWGN would lead to further variations in the received odor concentration during symbol detection, which could potentially induce errors. 

 \section{Selection of performance governing parameters}
 
  The performance of an OMC system in terms of the achievable capacity  depends on whether a transmitted symbol (encoded at an odor intensity level) is received within its corresponding acceptable odor concentration range or not. Moreover, the OMC system should also be able to meet up with the data rate requirements of any potential application avenue as well. Both these aspects largely depend on the choices of $v$ and $\tau$ that are used in the OMC system, and henceforth, it is fitting to call both of these variables as \textit{performance governing parameters}. The data rate required in an application is known beforehand and thus choice of $\tau$ can be controlled and set accordingly. We also lay emphasis on the fact that depending on the environment setting (bounded/unbounded) we could control $v$ as well to a certain extent. For instance, in a bounded environment where an artificially-directed constant airflow is present along the odor molecule propagation path from Tx to Rx, $v$ is a controlled parameter. However, in an unbounded environment where the constant streamline flow of the wind governs the advection of odor molecules, $v$ becomes uncontrollable. In this section, we provide numerical results, which lay emphasis on how the choices of airflow speed ($v$) and symbol period ($\tau$) governs the magnitude of deterioration in the transmitted odor concentration over space-time until it finally reaches the Rx. Hence, we are more concerned with the loss in concentration over space-time due to the choices of $v$ and $\tau$, we obtain the results by only considering the propagation noise caused by uncertainty in the diffusion coefficient $D$. We emphasize that role of AWGN noise only arise during the symbol detection phase at Rx and not in the symbol propagation phase because this is the noise that arises within the electronic hardware components of the receiver. Lastly, the results in this section are obtained using the following fixed quantities: $A=68.2320$ $ou/m^3$, i.e., we are transmitting the bit \enquote{1} symbol, $R=50$ $m$, $T=298$ \textit{K} at $1$ \textit{atm} of environment pressure.
 \subsection{Variation in Airflow Speed}
 First, we examine the effect of variation in $v$ for $v\in\{0.5,1,1.5,2,2.5,3,3.5,4\}$ $m/s$ on the received odor concentrations. These values of $v$ are considered, because it is fairly easy to generate them practically (in a bounded environment setting) and are easily found in the natural wind flow as well (in an unbounded environment setting). Much stronger airflows could be explored as well if necessary in the case of bounded environment. However, it requires, the system designer to also take into consideration the power resources that are available in the system (besides the OMC system). Hence, in this paper, we only focus on the aforementioned values of $v$ that would work well for both bounded and unbounded environment scenarios. For this analysis, we chose the symbol duration as $\tau=20$ $\mu s$, which represents a typical low data rate requirement application avenue. Now, we carry out the Monte-Carlo simulations on (\ref{Solution}) to obtain several readings for the received odor concentration for each $v$ value and then took their mean ($\mu$) to get an accurate understanding of received odor concentration for each such value of $v$. We present the obtained simulation results in Table \ref{v_Macroscale}. Furthermore, we include the expected standard deviation ($\sigma$) values for each $v$ as well to highlight the uncertainty in the received mean odor concentration readings. 
 \begin{table}[t!]
     \centering
      \caption{Effect of variations in $v$ and $\tau$ on Odor conc. at Rx }
     \begin{tabular}{|c|c||c|c|}
     \hline
         \multicolumn{2}{|c||}{\textbf{Variation in $v$ ($\tau=20$ $\mu s$)}} &  \multicolumn{2}{|c|}{\textbf{Variation in $\tau$ ($v=4$ $m/s$)}}    \\
          \hline
          \textit{Air speed}  & \textit{Conc : $\mu(\sigma)$} & \textit{Symbol period} & \textit{ Conc :$\mu(\sigma)$}\\
           \hline
         $v=0.5$ $m/s$ & 0(0) & $\tau=1$ $ns$  & 0(0) \\
          \hline
         $v=1$ $m/s$ & 3.5967(1.5541) & $\tau=1$ $\mu s$ & 0(0)  \\
          \hline
         $v=1.5$ $m/s$ & 39.4832(0.7017) & $\tau=2$ $\mu s$ & 27.8954(0.9884) \\
          \hline
         $v=2$ $m/s$ & 52.0722(0.3946) &  $\tau=5$ $\mu s$ & 52.0643(0.3895) \\
          \hline
         $v=2.5$ $m/s$ & 57.9014(0.2435) & $\tau=$10 $\mu s$ & 60.1434(0.1952) \\
          \hline
         $v=3$ $m/s$ & 61.0535(0.1734) & $\tau=20$ $\mu s$ & 64.1969(0.0956) \\
          \hline
         $v=3.5$ $m/s$ & 62.9527(0.1272) & $\tau=100$ $\mu s$ & 67.4241(0.0194) \\
          \hline
         $v=4$ $m/s$ & 64.1877(0.0976) & $\tau=1$ $m s$ & 68.1513(0.0020) \\
          \hline
     \end{tabular}
    
     \label{v_Macroscale}
     \vspace{-0.41cm}
 \end{table}
 From the Table \ref{v_Macroscale}, we interpret that firstly no odor molecule reaches the Rx in the presence of extremely low airflow speed ($v=0.5$ $m/s$), which was expected as the Tx-Rx separation here is of a macro-scale scenario rather than a nano-scale one. Next, when the airflow speed is slightly increased ($v=1$ $m/s$), then, we begin to observe odor reception but still the received odor concentration is far off from the value required for the correct detection of the symbol as in Table \ref{range_table}. The reason for this behavior is the fact that for such low air speeds the diffusion phenomena dominates over the advection phenomena and as a result emitted odor molecules tend to follow a much more random path rather than an advection induced guided path. This eventually results in rapid deterioration of initially emitted concentration over space-time ultimately leading to reception of extremely low levels of odor concentration at Rx.\\
 
 Furthermore, as we increase the directed airflow speed we observe that mean received odor concentration values also increases and it eventually enters the desired zone (for $v\geq 2$ $m/s$), which would lead to correct detection at Rx. Moreover, it is also observed that as $v$ increases the standard deviation ($\sigma$), and thus, in turn variance ($\sigma^2$) in the received odor concentration also decreases. This implies that the randomness associated with the received odor concentration of a symbol is reduced considerably provided we ensure the presence of a strong directed airflow. On this basis, it sounds convincing that ensuring a stronger directed airflow is appropriate as it would minimise the uncertainty in the received odor concentration and also achieve a mean received odor concentration that is well within the desired detection range. 

\subsection{Variation in symbol duration}
 The objective of this subsection is to form an understanding of the appropriate choice of the symbol duration $\tau$ and for this we implement Monte Carlo simulations whilst setting the airflow speed at $v=4$ $m/s$. We emphasize that in order to achieve a good performance, the OMC system should indeed incorporate a strong directed airflow as evident from Table \ref{range_table}. The results of the simulations carried out on (\ref{Solution}) for different $\tau$ values is also included in Table \ref{range_table}. Hence, it is evident that for extremely short symbol period when $\tau=1$ $ns$ or $\tau=1$ $\mu s$, there is no odor concentration that is being received which highlights that such low symbol periods (and in high data rate) could not be supported by the OMC system for this very choice of $v$. Nevertheless, as we increase $\tau$ to the units of $\mu s$ we observe that now the odor molecules are being received at Rx and that too with a significant rise in mean value of concentration when we adjusted the symbol period to $\tau=2$ $\mu$ $s$. For all the readings that follow, we observe a steady rise in the mean received odor concentration and a reduction in the associated variance as we were expecting. However, it is important to highlight that the achievable data rate in an OMC system is inversely proportional to the symbol time slot, i.e., $\propto \frac{1}{\tau}$ and for ideal binary communication systems this data rate is equal to $\frac{1}{\tau}$ itself. From Table \ref{v_Macroscale}, when $\tau=10$ $\mu s$, the mean received concentration is 60.1434 $ou/m^3$ and the max achievable data rate in ideal scenario is $100$ \textit{kbps}. Similarly for $\tau=1$ $ms$, the mean received concentration is 68.1513 $ou/m^3$ and the max achievable data rate is only $1$ \textit{kbps}. Although in both cases the received odor concentration is within the correct detection range, a higher $\tau$ would lead to lower achievable data rates and vice-versa and this analogy holds true for practical communication systems. Hence, depending upon the data rate requirements of an application avenue, the choice of $\tau$ for a particular $v$ would have to appropriately chosen. For a rough estimate the following thumb rule could be helpful while selecting appropriate $v$ for a given $\tau$, i.e.,
\begin{enumerate}[(i)]
    \item Low data rate required ($kbps$), i.e., $\tau$ ($\mu s$), lower $v$ will suffice.
    \item High data rate required ($Mbps$), i.e., $\tau$ ($ns$), higher $v$ will be required.
\end{enumerate}

When $v$ is extremely strong between Tx-Rx for an unbounded environment, the natural winds would become turbulent. Hence, (\ref{Solution}) would become invalid for such cases. However, such scenarios are very rare especially at low altitudes where the OMC system is more likely to be employed and hence are not covered in this paper. If required, they can be analysed by directly expanding (\ref{tracer}) directly along the spacial dimensions considered in the system model and also considering variations in eddy diffusivity coefficient $K$ over space. 

 \section{Channel Capacity of Odor Molecular Communications } 
 
\begin{figure}[t!]
     \centering
     \includegraphics[width=3.4in,height=1.8in]{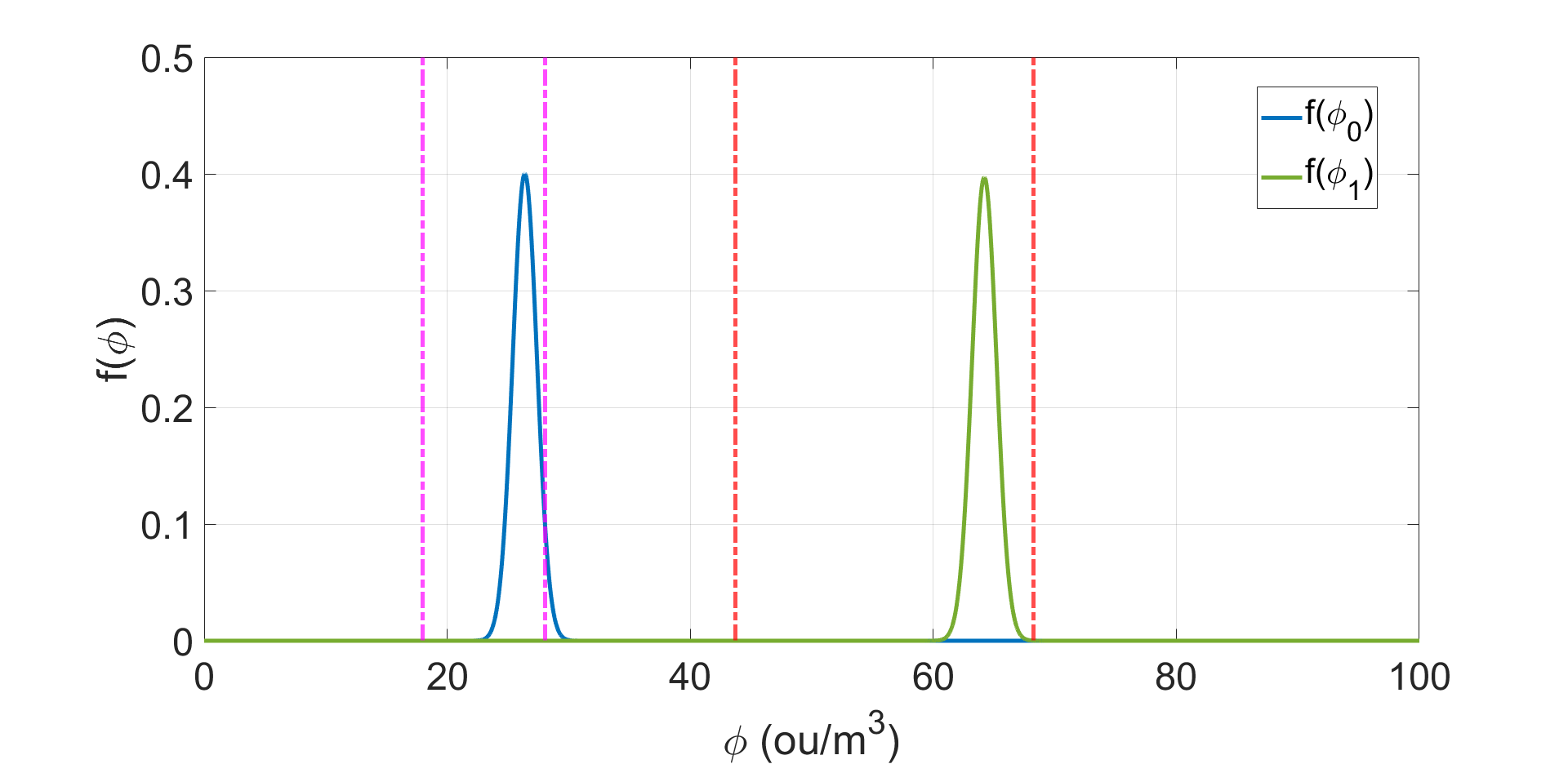}
     \caption{Gaussian probability density functions from (\ref{pdf0}) and (\ref{pdf1})}
     \label{gauss_pdfs}
 \end{figure}

 In this section, we present the methodology that has to be undertaken to determine the channel capacity for an OMC system and we commence by defining the fixed entities that we have used for its evaluation. They are as follows:
 \begin{itemize}
    \item Odorant : Benzene at $T=298$ \textit{K} at $1$ \textit{atm} 
    \item Diffusion coefficient : $D\in[9.078\times 10^{-6},9.868\times 10^{-6}$] $m^2/s$.
     \item Tx-Rx separation : $R=50$ $m$
     \item Externally directed airflow speed : $v=4$ $m/s$
     \item Symbol slot duration : $\tau=20$ $\mu s$.
     \item AWGN at Rx with zero mean ($\mu_n=0$) and unit variance ($\sigma^2_n=1$).
 \end{itemize}
 Recall that the $D$ cannot be an exact value, which makes $D$ the only parameter in OMC system, which is indeed a random variable as emphasized on earlier.
 Now, we present a framework, which provides an insight of the procedure that has to be undertaken in order to determine the channel capacity of an OMC system.   
 \begin{enumerate}[1)]
     
    \item Here, we use the $M = 2$ symbols ($\{S_0 = 0 , S_1 = 1\}$) OISK modulation to encode the information, implying that both the symbols will be encoded at different odor intensities (in turn concentrations). 
    
    \item Let $\phi_0$ and $\phi_1$ be the random variables denoting the received concentrations of the odor at Rx when symbols
    \enquote{0} and \enquote{1} are sent, respectively.
    
    \item We describe both these random variables using the Gaussian probability distribution function for accurately describing their behaviours. Whilst evaluating the mean and variance for both of these random variables we also incorporate adverse effects such as propagation noise and the AWGN in the OMC system and then implement the Monte-Carlo simulations.

    \item Let $X$ and $Y$ be the random variables denoting the transmitted symbol and the received symbol, respectively. Then, the mutual information between these random variables can be defined as \cite{BP_lathi}
    \vspace{-0.2cm}
    \begin{multline}
       M(X; Y) = \sum_{i=0}^{1} \sum_{j=0}^{1} Pr(X_i)Pr(Y_j|X_i) \\
       \times \log_2 \bigg(\frac{Pr(Y_j|X_i)}{\sum_{i=0}^{1} Pr(X_i)Pr(Y_j|X_i)} \bigg)
       \label{mutual}
    \end{multline}
    where $X_i$ and $Y_j$ both $\in\{S_0 = 0 , S_1 = 1\}$ 

    \item For a single symbol time slot, the capacity is the maximisation of mutual information over the all possible Tx symbol probabilities, i.e.,
    \vspace{-0.2cm}
    \[
        C = \max_{Pr(X_i)}  M(X; Y) 
    \]

    where $P(X_i)=\{p_0,p_1\}$ which are the probabilities of the Tx symbol being \enquote{0} and \enquote{1}, respectively. Now, say if $p_0=\alpha$ then surely $p_1=1-\alpha$. Hence, in other words channel capacity is actually the maximisation of the mutual information over the Tx symbol probability parameter $\alpha$.
    \begin{equation}
        C = \max_{\alpha}  M(X; Y) 
        \label{capacity}
    \end{equation}
    where the input symbol probability variable $\alpha\in[0,1]$.
\end{enumerate}
This completes the framework of the channel capacity determination.\\
In order to obtain a numerical value for the channel capacity we would be using (\ref{mutual}) and so we first need to evaluate all the conditional probability expressions that are used in (\ref{mutual}). We begin by defining all the necessary conditional probability expressions in terms of the received odor concentration as follows:
\begin{enumerate}[(a)]
    \item $Pr(Y_0|X_0) = Pr(17.9815\leq\phi_0\leq28.0463)$
    \item $Pr(Y_0|X_1) = Pr(17.9815\leq<\phi_1\leq28.0463)$
    \item $Pr(Y_1|X_0) = Pr(43.7448\leq\phi_0\leq68.2320)$
    \item $Pr(Y_1|X_1) = Pr(43.7448\leq\phi_1\leq68.2320)$
\end{enumerate}
These limits have been chosen in accordance with Table \ref{range_table}.
To evaluate them, we now define the random variables $\phi_0$ and $\phi_1$ as the gaussian random variables having probability density functions $f(\phi_0)$ and $f(\phi_1)$, respectively, i.e.,
\vspace{-0.1cm}
\begin{subequations}\label{pdfs}
\begin{equation}\label{pdf0}
    f(\phi_0) = \frac{1}{\sigma_0\sqrt{2\pi}} exp\bigg(\dfrac{-(\phi_0-\mu_0)^2}{2\sigma_0^2}\bigg)
   \end{equation}
   \text{and}
\begin{equation}\label{pdf1}
   f(\phi_1) = \frac{1}{\sigma_1\sqrt{2\pi}} exp\bigg(\dfrac{-(\phi_1-\mu_1)^2}{2\sigma_1^2}\bigg)
   \end{equation}
\end{subequations}
where $\mu_0$, $\sigma_0$ and $\mu_1$, $\sigma_1$ are the mean and standard deviation of the random variables, $\phi_0$ and $\phi_1$, respectively.\\
 Next, we incorporate the adverse effects of propagation noise (by taking $D$ as a random variable) along with the AWGN noise (which is present at the Rx) on the transmitted odor symbols in the Monte-Carlo simulations. These simulations where carried out (for both the symbols \enquote{0} and \enquote{1}) using the MATLAB software in order to determine the mean and variance in the received odor concentrations. The AWGN noise used in this simulation has zero mean, i.e., $\mu_n=0$ and unit variance, i.e., $\sigma^2_n=1$.  Finally, for the aforementioned OMC system parameters along with propagation noise and AWGN noise, the mean ($\mu$) and standard deviation ($\sigma$) for the random variables $\phi_0$ and $\phi_1$ that were obtained from the simulations are $\mu_0=26.3879$, $\sigma_0 =$ 0.9972 and $\mu_1=$ 64.2000, $\sigma_1 =$ 1.0049, respectively.
By substituting these values back in (\ref{pdfs}), we can straightforwardly obtain a graphical view of these pdfs as shown in Fig. \ref{gauss_pdfs}, where the magenta and red dashed lines denotes the appropriate concentration ranges for both random variables respectively as in Table \ref{range_table}.

 With the probability density functions in place, we now proceed with the evaluation of the necessary conditional probability expressions. We have incorporated some mathematical notations to simplify the closed form expression of mutual information and in turn of channel capacity as well. Let $Pr(Y_0|X_0)=P_{00}$, $Pr(Y_0|X_1)=P_{01}$, $Pr(Y_1|X_0)=P_{10}$ and $Pr(Y_1|X_1)=P_{11}$. Mathematically they are computed as follows:
 \begin{subequations}\label{conditional_prob_expressions}
\begin{equation}\label{p_00}
    P_{00} = \int_{17.9815}^{28.0463} f(\phi_0) d\phi_0
   \end{equation}
\begin{equation}\label{p_01}
     P_{01} = \int_{17.9815}^{28.0463} f(\phi_1) d\phi_1
   \end{equation}
\begin{equation}\label{p_10}
     P_{10} = \int_{43.7448}^{68.2320} f(\phi_0) d\phi_0
   \end{equation}
\begin{equation}\label{p_11}
     P_{11} = \int_{43.7448}^{68.2320} f(\phi_1) d\phi_1
   \end{equation}
\end{subequations}
 where $f(\phi_0)$ and $f(\phi_1)$ are as defined in (\ref{pdfs}). Specifically, using $\mu_0=26.3879$, $\sigma_0 =$ 0.9972, $\mu_1=$ 64.2000 and $\sigma_1 =$ 1.0049, we obtain the following values for the conditional probabilities: $P_{00}=0.951849$, $P_{01}=0$, $P_{10}=0$ and $P_{11}=0.99997$.
 Now, we expand (\ref{mutual}) using the aforementioned terminology of conditional probabilities to obtain a generic, closed form analytical expression for the mutual information $M(X;Y)$, i.e.,
\vspace{-0.1cm}
 \begin{multline} 
     M(X;Y) = \alpha P_{00} \times  \log \bigg(\frac{P_{00}}{\alpha P_{00} + (1-\alpha)P_{01}}\bigg)  \\+ \alpha P_{10} \times  \log \bigg(\frac{P_{10}}{\alpha P_{10} + (1-\alpha)P_{11}}\bigg)  \\ + (1-\alpha) P_{01} \times  \log \bigg(\frac{P_{01}}{\alpha P_{00} + (1-\alpha)P_{01}}\bigg) \\+ (1-\alpha) P_{11} \times  \log \bigg(\frac{P_{11}}{\alpha P_{10} + (1-\alpha)P_{11}}\bigg)  
     \label{generic_mutual}
 \end{multline}

\begin{figure}[t!]
     \centering
     \includegraphics[width=3.4in,height=1.8in]{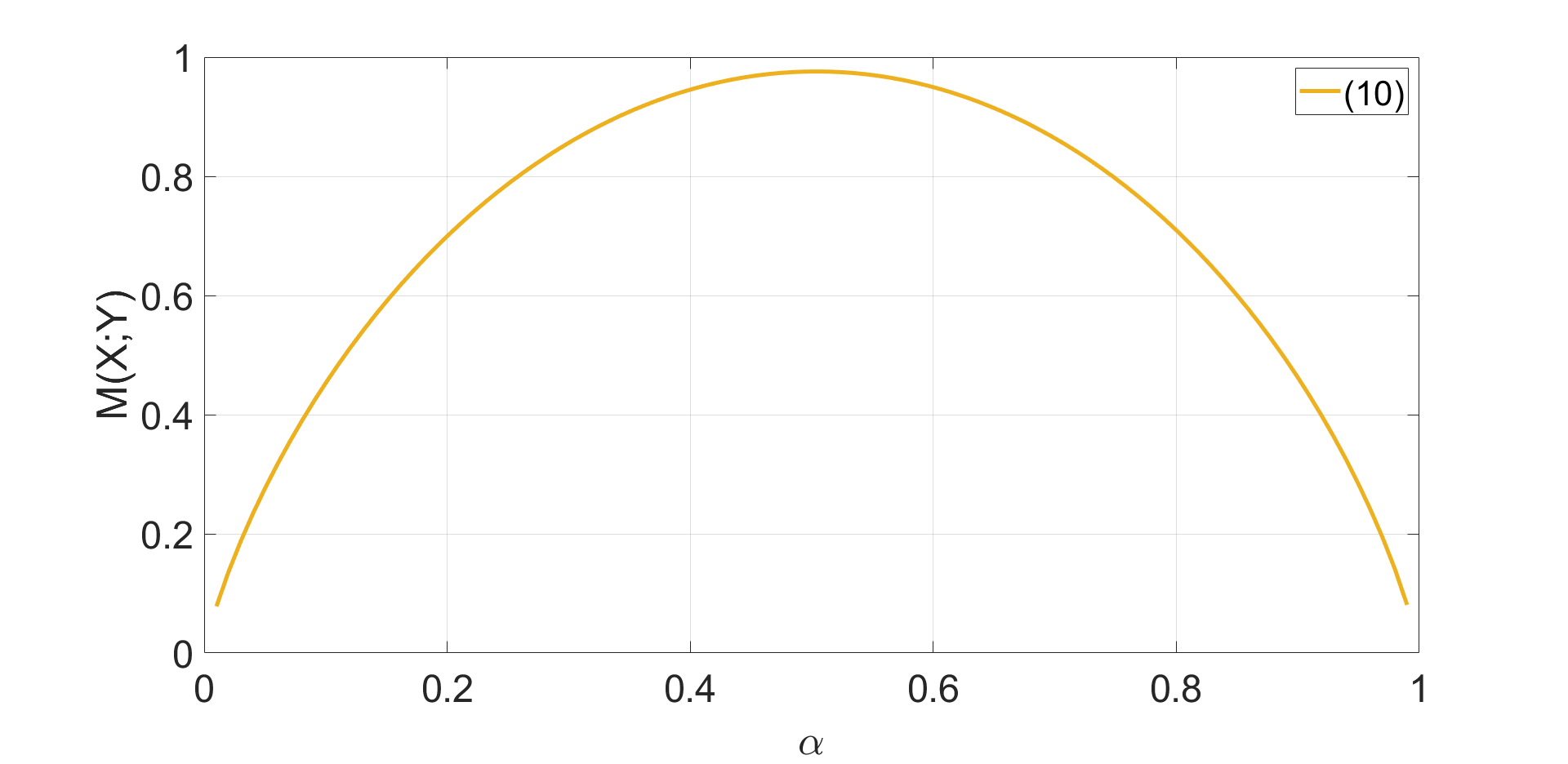}
     \caption{Variation in mutual information with input symbol probability}
     \label{mutual_basic}
 \end{figure}

 Upon the substitution of the calculated conditional probabilities $P_{00}$, $P_{01}$, $P_{10}$ and $P_{11}$ values into (\ref{generic_mutual}), it reduces down to yield a compact analytical expression for the mutual information as
 \vspace{-0.2cm}
 \begin{multline}
     M(X;Y) =  0.951849 \times \alpha  \log_2 \bigg(\frac{1}{\alpha}\bigg) + \\ 
     0.99997 \times (1-\alpha) \log_2 \bigg( \frac{1}{1-\alpha}\bigg)
     \label{final_mutual}
 \end{multline}
 The variation of the mutual information with the input symbol probability, i.e., $\alpha$ is shown in Fig. \ref{mutual_basic}. Finally, using (\ref{capacity}) we define a closed-form analytical expression for the channel capacity for a single symbol time slot as:
 \vspace{-0.3cm}
 \begin{multline}
     C = \max_{\alpha} \Bigg[ 0.951849 \times \alpha  \log_2 \bigg(\frac{1}{\alpha}\bigg) + \\ 
     0.99997 \times (1-\alpha) \log_2 \bigg( \frac{1}{1-\alpha}\bigg) \Bigg]
     \label{final_capacity}
 \end{multline}
 Here, $C$ has the unit of \textit{bit/slot}. Now, the maximisation of $M(X;Y)$ occurs at $\alpha=0.5051$ which implies that the channel capacity is achieved whenever the occurrence of the two symbols is almost equally probable in the transmitted information. The numerical value of the capacity was henceforth found to be $C=0.9759$ \textit{bit/slot}, that too in the presence of both the noise based adverse effects. However, we are not yet able to fully utilize the channel as the ideal value of the channel capacity for $M=2$ scheme used here is $1$ \textit{bit/slot}. As a result with a symbol period of $\tau=20$ $\mu s$ we can transmit a max of $0.9759$ bits without error. This implies that the maximum achievable data rate for this scenario is $48.795$ \textit{kbps} ($50$ \textit{kbps} ideally), which indicates that there is certainly a room for further improvement, which we do via the means of channel capacity analysis carried out in Section VI. Nevertheless, achieving this data rate would make the OMC system very suitable for being used in applications involving low data rates. \\
 For we now work towards providing a generic closed form expression that would evaluate the channel capacity of the OMC system. It is to be observed that as the intensity bandwidth of each intensity level is $0.5$ the corresponding concentration ranges are quite wide. Moreover, there is more than enough separation between the concentration ranges of both the symbols. Hence, unless the AWGN induced disturbance is extremely large ($\sigma_n>5$) the pdfs of the random variables i.e. $f(\phi_0)$ and $f(\phi_1)$ are not going to cross-over into each others range. Thus, the conditional probabilities corresponding to cross-detection of symbols viz $P_{01}$ and $P_{10}$ are  equal to zero. Thus, (\ref{final_capacity}) can be extended to yield a generic compact closed form analytical expression for the channel capacity, i.e.,
 \begin{equation}
     C = \max_{\alpha} \Bigg[ P_{00} \times \alpha  \log_2 \bigg(\frac{1}{\alpha}\bigg) + 
     P_{11} \times (1-\alpha) \log_2 \bigg( \frac{1}{1-\alpha}\bigg) \Bigg]
     \label{generic_capacity}
 \end{equation}
 The usage of (\ref{generic_capacity}) simplifies and improves the efficiency of capacity evaluation on mathematical software like MATLAB. We use (\ref{generic_capacity}) throughout the remainder of the paper where we delve into the analysis of channel capacity under the effects of variation in airflow speed, temperature of environment and AWGN at Rx and also aim to achieve the ideal capacity in the OMC system.

 \section{Analyzing the Effects of Airflow Speed, Temperature and AWGN on the Capacity}
 \vspace{0.2cm}
In this section, we analyse the effects of variation in parameters like airflow speed ($v$), temperature ($T$) of environment and the AWGN noise power ($\sigma_n^2$) on the max achievable channel capacity on a case-by-case basis. Furthermore, we also utilise the results from this analysis to determine a suitable value for $v$ which would be required to ensure that the achievable channel capacity of the system is almost equal to the ideal capacity of $1$ \textit{bit/slot}. Here, we do not explicitly analyse the effect of $\tau$ on the capacity because the value of $\tau$ that is chosen for the OMC system depends on the data rate that is required by an application avenue where the OMC system is being applied. For the analysis in this section we keep $\tau=20$ $\mu s$, which implies a typical low data rate scenario. It is to be noted that variation in $T$ directly affects the diffusion coefficient of an odorant, and hence, they would alter the effect of propagation noise. Table \ref{simulationparamter_table} lists the values of the standard OMC system parameters that were used to obtain the simulation results on channel capacity using the benzene odorant at \textit{1 atm} environment pressure. Unless stated otherwise, these parameters will have the values as in Table \ref{simulationparamter_table} 
 \begin{table}[t!]
     \centering
     \caption{Simulation Parameters}
     \label{simulationparamter_table}
     \scalebox{1.1}{
     \begin{tabular}{|c|c|}
        \hline
         \textbf{Parameter} & \textbf{Value} \\
         \hline
         \hline
         \textit{Temperature:} $T$ & \textit{298} $K$ \\
         \hline
         \textit{Diffusion coefficient:} $D$ & $D\in[9.078,9.868]\times 10^{-6}$ $m^2/s$ \\
         \hline
         \textit{Tx-Rx separation:} $R$ & \textit{50} $m$ \\
         \hline
         \textit{Airflow speed:} $v$ & \textit{4} $m/s$ \\
         \hline
         \textit{Symbol slot duration:} $\tau$ & \textit{20} $\mu s$\\
         \hline
         \textit{AWGN mean and variance} & $\mu_n=0$ , $\sigma^2_n=1$ \\
         \hline
     \end{tabular}}
     
 \end{table}
 \newline

 \subsection{Effect of $v$ on $C$}

\begin{figure}[t!]
     \centering
     \includegraphics[width=3.4in,height=1.8in]{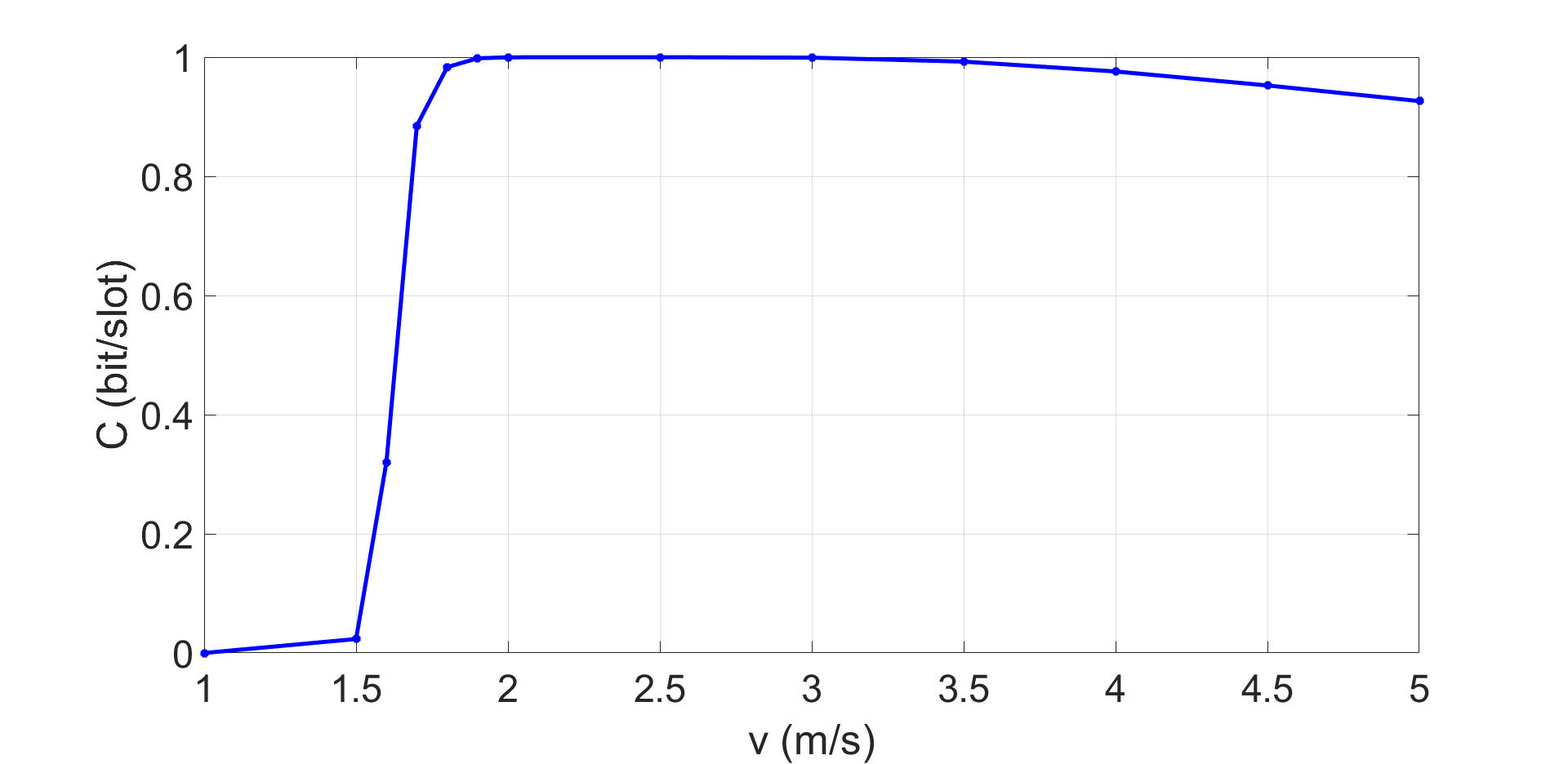}
     \caption{Variation in capacity with airflow speed}
     \vspace{-0.4cm}
     \label{Capacity_v}
 \end{figure}

 Fig. \ref{Capacity_v} illustrates the effect that the airflow speed has on the channel capacity. Initially, for significantly low airflow ($v<1.5$ \textit{m/s}) the channel capacity is extremely low, which is expected as here diffusion induced randomised motion of odor molecule is dominating over the guided flow provided by advection. Nevertheless, once we go beyond this value i.e. $v>1.5$ \textit{m/s} a significant rise in the capacity value is observed. The capacity values now are extremely close to the  ideal value of $1$ \textit{bit/slot} that we want to achieve in the system. For $v=2.5$ \textit{m/s} the evaluated capacity reaches it maximum of $0.9999\approx 1$ \textit{bit/slot}. However, afterwards when we further increase the airflow we witness that the capacity begins to dwindle. Since an increase in airflow speed implies domination of advection phenomenon apparently it may seem that this behavior is highly out of order. Nevertheless, there is an underlying factor that must be paid attention to. It is the mean received odor concentration value at Rx. From Table \ref{v_Macroscale}, it is evident that in presence of strong airflow, the received odor concentration is quite close to the initial concentration value. This initial concentration is also the boundary of the correct detection range. Due to the AWGN noise at Rx, the received odor concentration is likely to cross the correct detection boundary. This leads to reduction in the achievable channel capacity. This behavior is clearly evident from Fig. \ref{gauss_pdfs} whereby the pdf $f(\phi_0)$ is exceeding the correct detection boundary in the presence of a strong airflow of $v=4$ \textit{m/s}. Hence, it is appropriate to conclude that for the OMC system parameters listed in Table \ref{simulationparamter_table} we just require a fairly moderate airflow speed to ensure the maximum possible capacity in the channel. The maximum achievable capacity in this analysis was found to be $\approx 1$ \textit{bit/slot} at $v=2.5$ \textit{m/s} and as $\tau=20$ $\mu s$ it would correspond to a max achievable data rate of $50$ \textit{kbps} which pertains to low data rate. Furthermore, if this system is being operated in a bounded environment then also in order to ensure a high channel capacity we need to artificially provide only a fairly moderate airflow, which could be a low energy task. If its achievable then the proposed OMC system would be an energy efficient solution to obtain lower data rates.

\subsection{Effect of $T$ on $C$}

When the OMC system environment temperature experiences a change, the diffusion coefficient ($D$) gets directly affected by it. As seen from (\ref{Solution}), we can clearly see that the variation in odor molecule concentration over space-time is heavily dependent on the parameter $D$ as well and thus any change in $T$ is bound to have an effect on the received odor concentration as well. Fuller's method as described in \cite{Diffusion_coeff} provides us with an analytical expression via, which we evaluate the Diffusion coefficient $D$ at \textit{298 K} by extrapolating the value of $D$ that has been measured at any other $T$ in kelvin. Mathematically,
\[
 D(298) = D(T) \Big(\dfrac{298}{T}\Big)^{1.75}
\]
From this expression we can also back calculate the Diffusion coefficient values at any arbitrary temperature, i.e., $D(T)$ if we know the value of $D(298)$. Thus, from the perspective of our requirement we can alternately express Fuller's method as:
\begin{equation}
    D(T) = D(298) \Big(\dfrac{T}{298}\Big)^{1.75}
    \label{Fuller_formula}
\end{equation}
Moreover, in \cite{Diffusion_coeff} it has been emphasized that the results obtained from this method for the benzene odorant are in good agreement with the corresponding practically measured diffusion coefficient values of benzene \cite{Nagata1970},\cite{Lee_Wilke1954},\cite{Cohen1960},\cite{Lugg1968} as well, especially over the temperature range of $T\in[290,400]K$. Generally, for both bounded/unbounded OMC system environments it is reasonable to expect the environment temperature of an OMC system to be well within the aforementioned temperature range, and hence, we center our channel capacity analysis over the temperature range of $T\in[290,400]K$. In section VI.A, we observe that at room temperature ($T=298$ \textit{K}), we obtain the maximum channel capacity of $\approx 1$ \textit{bit/slot} for $v=2.5$ \textit{m/s}, and hence, to study the effect of variation of $T$ on the capacity it is best to keep $v=2.5$ \textit{m/s} in the OMC system. All other system parameters are as in Table \ref{simulationparamter_table}.

\begin{figure}[t!]
    \centering
    \subfigure[\label{Capacity_T}]{\includegraphics[width=3.4in,height=1.8in]{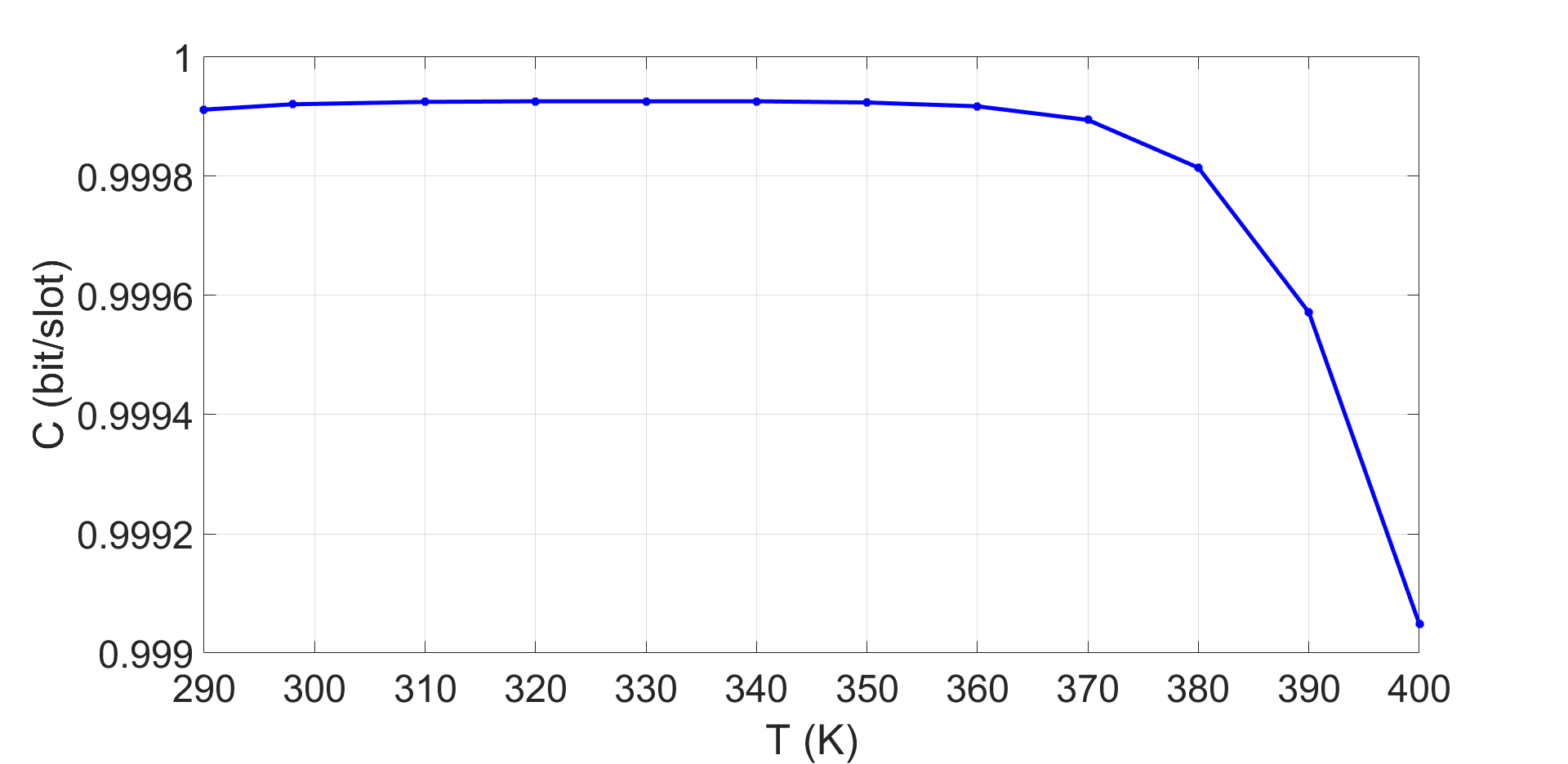}}
    \subfigure[\label{conc_T}]{\includegraphics[width=3.4in,height=1.8in]{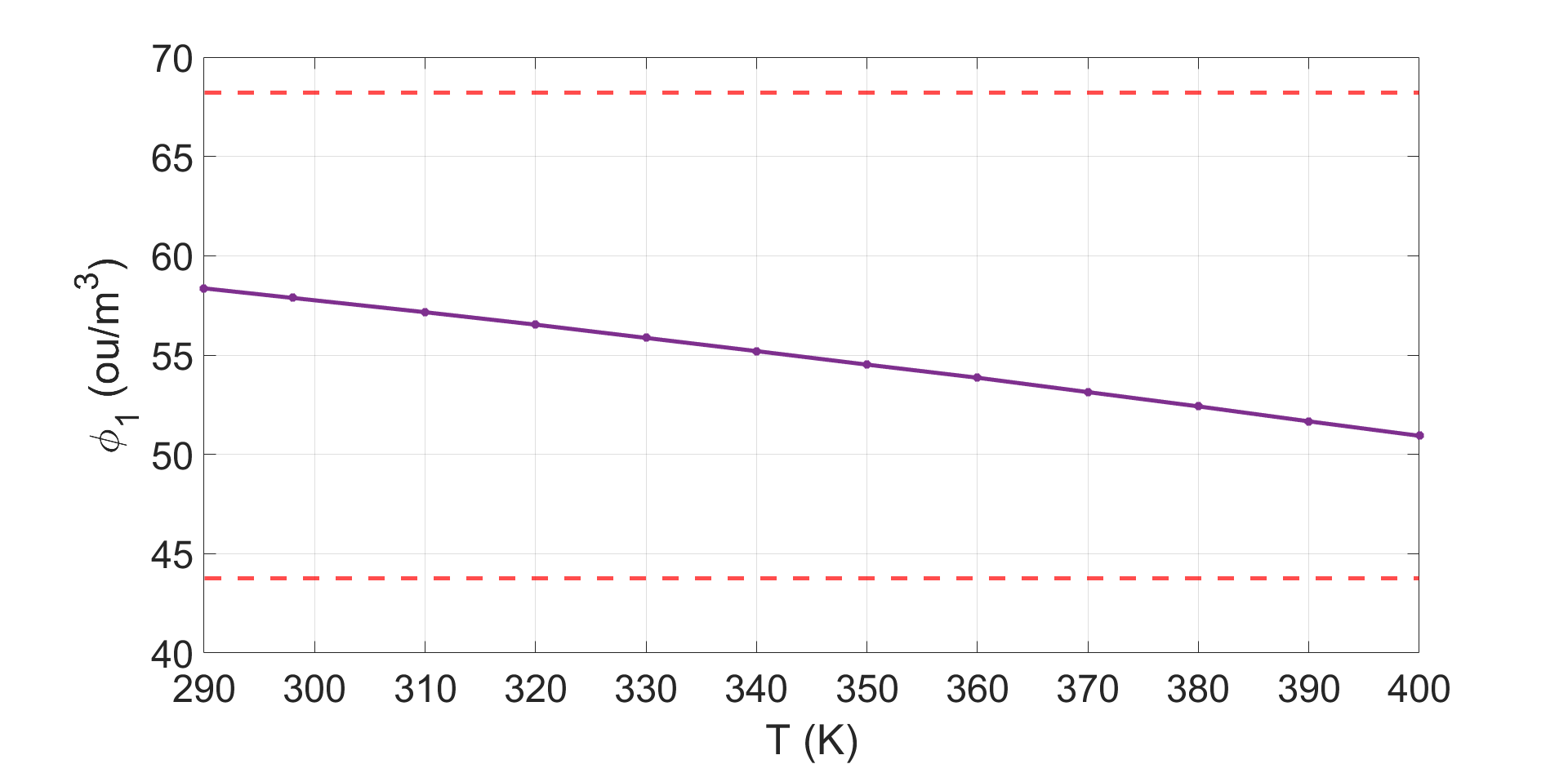}}
    \caption{a) Variation in capacity with $T(K)$ , b) variation in $\phi_1$ with $T(K)$}
    \vspace{-0.4cm}
    \label{Temperature_figures}
\end{figure}
 Using (\ref{Fuller_formula}) for each temperature value $T$ its equivalent new diffusion coefficient $D(T)$ is determined. Afterwards, Monte Carlo simulations are implemented to obtain the the mean and standard deviations for the received odor concentration random variables namely $\phi_0$ and $\phi_1$ following, which the capacity was obtained as demonstrated in Section V. Fig. \ref{Capacity_T} illustrates the effect of temperature on the achievable channel capacity. It is evident that for the majority of the temperature range under consideration the capacity is extremely close to its ideal value of \textit{1 bit/slot}. Only at very high environment temperature ($T>370$\textit{K}) we observe that capacity begins to decline. However, the decline is not very drastic as the capacity is still quite high for instance even at $T=400$\textit{K} we are able to achieve a capacity of $C=0.999048$ \textit{bit/slot}. On the basis of this data, we emphasize that the channel capacity of the proposed OMC system is resistant to changes in the environment temperature. This is beneficial to ensure optimal performance of the system in those application avenues which may not have a constant temperature. The reason due to which the proposed OMC system is able to withstand such variations in the temperature lies in the choice of intensity bandwidth that was used to encode an odor symbol within a odor intensity level. As seen from Fig. \ref{Weber_bands}, the choice of intensity bandwidth ($0.5$ here) that is used to represent an intensity level ($I$) directly controls the range of odor concentration that corresponds to correct detection of an odor symbol. Wider the intensity bandwidth, broader would be the corresponding concentration range and better would be the resistance to temperature changes. Fig. \ref{conc_T} is shown to further illustrate this point. This figure illustrates the effect of temperature on the mean value of random variable $\phi_1$. As the temperature of environment increases, the thermal energy of propagating odor molecules also increases inducing even more random motion in the odor molecules, leading to a steady decline in $\phi_1$ as we expect. However, intensity bandwidth being $0.5$ here ensures that the corresponding concentration range for correct detection is wide enough to keep the declining received odor concentration well within the desired concentration range itself. The concentration range for correct symbol detection is highlighted by red dashed lines in Fig . \ref{conc_T}. 
Hence, the choice of intensity bandwidth ultimately makes the achievable channel capacity of proposed OMC system resilient to temperature variations in the system environment, which further solidifies the impact of our proposed work. 

\subsection{Effect of AWGN on $C$}
\begin{figure}[t!]
     \centering
     \includegraphics[width=3.4in,height=1.8in]{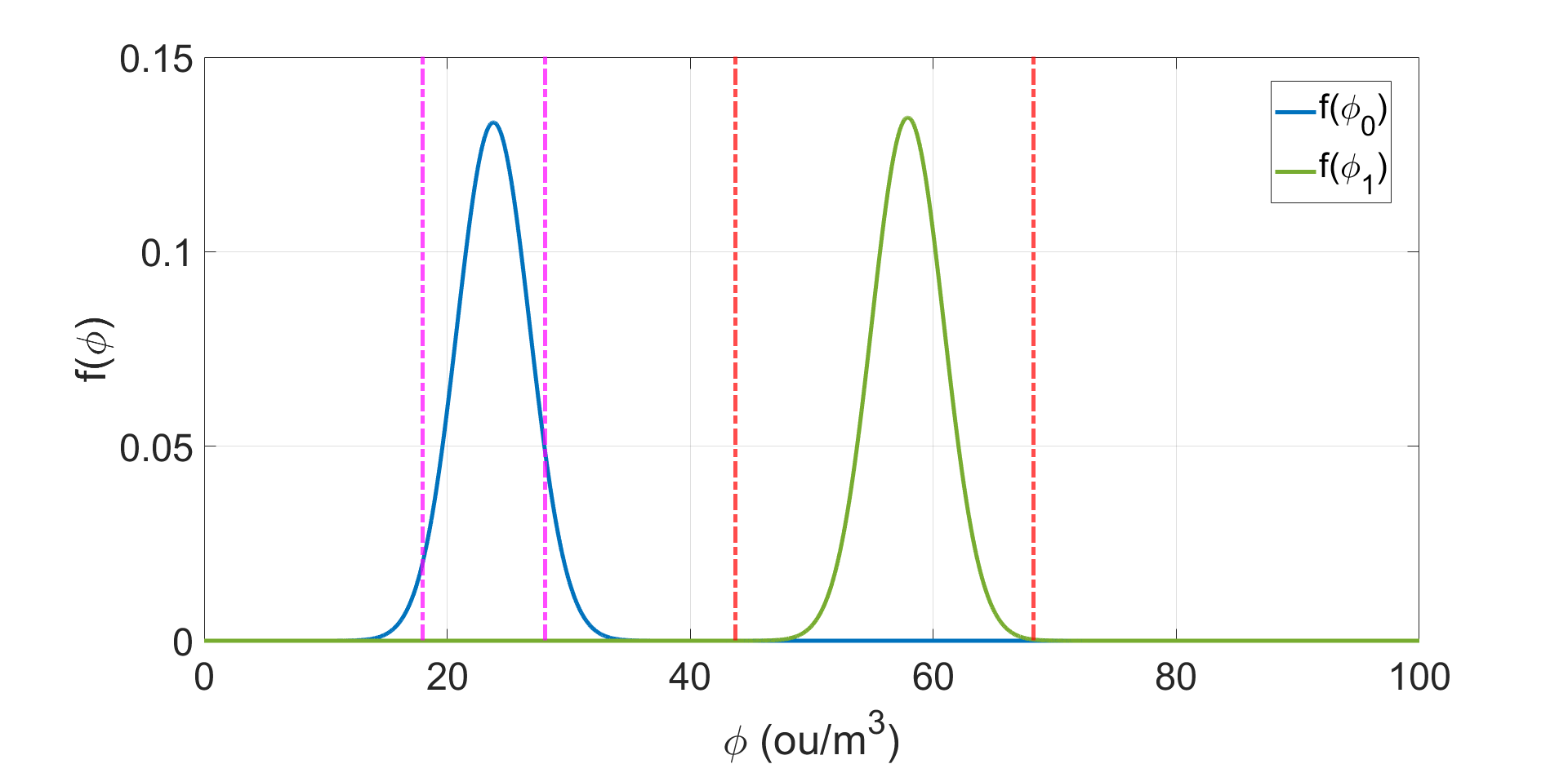}
     \caption{Affect of AWGN on (\ref{pdfs})}
     \label{AWGN_pdf}
 \end{figure}

\begin{figure}[t!]
     \centering
     \includegraphics[width=3.4in,height=1.8in]{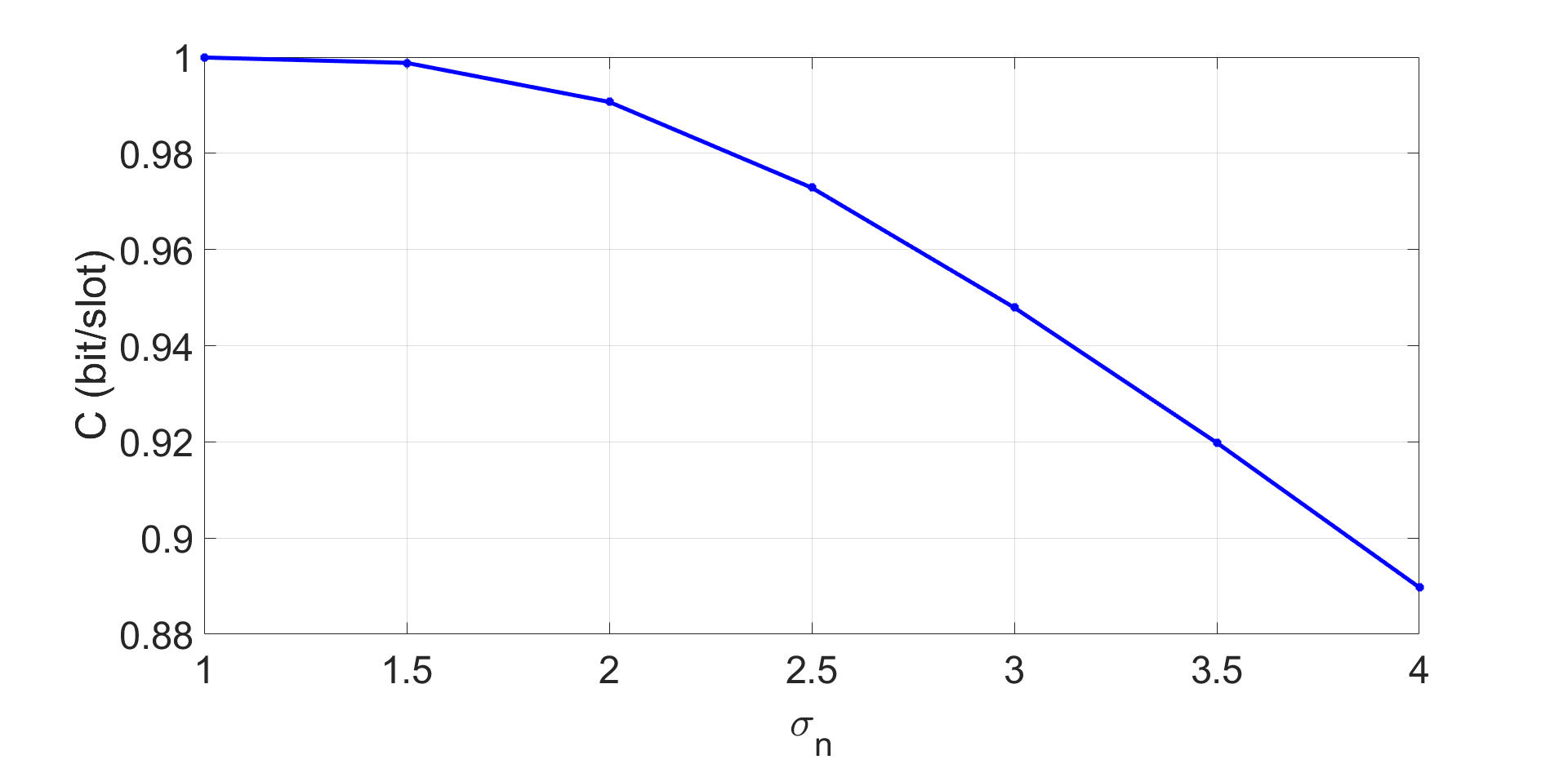}
     \caption{Variation in capacity with AWGN }
     \vspace{-0.4cm}
     \label{Capacity_AWGN}
 \end{figure}

In continuation with the capacity analysis in the previous subsection, we maintain the airflow speed as $v=2.5$ \textit{m/s}. All other OMC system parameters have been taken as in Table \ref{simulationparamter_table}. The AWGN has an adverse effect on the OMC system as this noise induces variations in the received odor concentration at Rx, which can cause the concentration value to cross its designated detection zone that corresponds to a correct symbol detection. Hence, it is imperative that the effect of variation in the AWGN power on the achievable capacity is analysed to understand how the OMC system performs as the influence of this noise gets more prominent. The AWGN noise has a unique characteristic that its mean is typically equal to zero, i.e., $\mu_n=0$ and its power is equivalent to the square of its standard deviation ($\sigma_n$), i.e., variance ($\sigma^2_n$) itself. Thus, the AWGN noise power is $P_n=\sigma_n^2$. While implementing the simulations, we keep the noise mean value fixed to zero as before and now we vary the noise standard deviation over the range of $\sigma_n\in[1,4]$. This also implies that we are indeed varying the AWGN noise power $P_n$. An increase in the AWGN noise power at the receiver leads to an increase in the variance of the random variables $\phi_0$ and $\phi_1$, which now causes substantial broadening of their respective gaussian pdfs as defined by (\ref{pdfs}). This is illustrated in Fig. \ref{AWGN_pdf}, where an AWGN of $\mu_n=0$, $\sigma_n=3$ has been considered. 
The gaussian pdfs in Fig. \ref{AWGN_pdf} are considerably broader (than those in Fig. \ref{gauss_pdfs}). This leads to a loss in the achievable capacity (down to $0.95$ \textit{bit/slot}). However, both these pdfs are still considerably within their respective concentration range of correct detection as indicated by the magenta and red colored dashed lines of Fig. \ref{AWGN_pdf}. This is due to choice of intensity bandwidth (0.5 here) which ensures sufficient width of the correct detection range. In this way, the proposed OMC system is able to combat AWGN induced disturbance. Moreover, it is evident from Fig. \ref{Capacity_AWGN}, that decline in capacity with an increase in standard deviation of AWGN is quite gradual.

 Lastly, via performing the channel capacity analysis we have been able to show that the ideal capacity per slot, i.e., $1$ \textit{bit/slot} can be achieved in the OMC system with airflow speed taken as $2.5$ \textit{m/s} with other parameters as in Table \ref{simulationparamter_table} . Furthermore, since the symbol time slot is taken as $\tau=20$ $\mu s$, the max achievable bit rate is $50$ \textit{kbps}, which is a typical low rate value. For different data rate requirements firstly the parameter $\tau$ has to be readjusted appropriately and channel capacity analysis needs to be performed to determine a suitable value for the airflow speed $v$. Furthermore, the system performance in challenging scenarios having temperature fluctuations and uncertain AWGN noise could also be observed for those cases.

\section{Conclusion and Future Work}
In this paper, we propose a novel OMC system model based on the OISK modulation scheme for conveying information, advection-diffusion mechanism for odor propagation, to obtain and analyse the channel capacity in artificial olfactory communication networks. Adverse effects like propagation noise and AWGN were incorporated in the proposed system to get a comprehensive understanding of the achievable capacity in realistic scenarios. Robustness of the system in circumstances, where $v$, $T$ and $\sigma_n$ undergo substantial changes has been highlighted with numerical and graphical results. Furthermore, the proposed OMC system model could also be potentially used as a base model and with further modifications it could also enable research of other associated ICT aspects as well. In this paper, the channel capacity and the corresponding capacity analysis yields results that particularly suits the low data rate requirement systems with a considerably wide Tx-Rx separation which makes the proposed work suitable for application in several Internet of Things (IoT) and Internet of Everything (IoE) based applications. Some of these are as follows:
\begin{itemize}
    \item Internet of Manufacturing Things: IoT sensors are often deployed in applications that often involve low data rate requirements. For instance, IoT sensors are used in a variety of environmental scenarios that may be either bounded or unbounded, to monitor/measure entities like temperature, humidity. These sensors communicate their measured data using low data rates. The proposed work is suitable for application here as it requires low natural wind speeds (unbounded) or low directed airflow (bounded) both which are easily found in nature or can be achieved artificially respectively.
    \item Internet of Agricultural Things: In modern agricultural practices artificial sensors are often deployed in the field to collect data about soil moisture, crop health, and even the weather conditions in the vicinity of the sensor. This data is transmitted back at low data rates via an ad-hoc network all the way back to the main computer node which further analyses the measured values. The implementation of the proposed OMC system in such applications is perfect, as besides having the capability of achieving a low data rate, the OMC system is also resistant to loss in achievable channel capacity due to temperature variations. This is a monumental advantage as the temperature in an agricultural field would always vary throughout the course of day/night. 
    \item Internet of Home Automation: Smart homes are composed of various smart IoE devices that carry out their tasks without requiring human intervention. Such autonomous devices include intelligent interior lightening systems, thermostats, smart security systems, automated motion detectors etc, all of which communicate with low data rates and their hardware setup includes various electronic components which can potentially induce noise related errors (via thermal noise) in the communication data. The application of the proposed work in this sector would ensure reliable low data rate communication along with making the systems channel capacity resilient to noise induced errors.
\end{itemize}
The channel capacity in the high data requirement scenarios with $M=2$ OISK scheme can be analysed using the same methodology as presented in sections IV, V and VI. To further increase the data rate $M=4$ (4 symbols encoded in 4 different intensity levels) OISK scheme could be used as well, which would in turn provide a max achievable capacity of $2$ \textit{bit/slot}. Besides the data rate centered industrial aspects, artificial olfactory systems can also find applicability in other avenues as well. Some of the very promising avenues include medical diagnosis \cite{NAUDIN_2014},\cite{Suh_2020}, healthcare treatment \cite{Mustapha_2020}, developing security systems and smart surveillance \cite{Troisi_2022}, inspection of food quality \cite{Dilara_unpublish} and advanced defence technology \cite{MITthesis}.

 In addition to the work presented in this paper, substantial research is required in the field of odor based molecular communications for their practical realisation. Here, we discuss some of the essential future work areas in the domain of artificial olfactory communications. In order to achieve the data rate values presented in this study, further research is required to develop novel state-of-the-art Tx designs which could execute odor symbol emissions at specific intensity levels within such narrow symbol time slots. Furthermore, novel Rx designs are also required that besides odor identification and detection could also distinguish between the different odor intensity levels possessed by the same odorant which is crucial for ensuring correct decoding of the received odor symbols. Moreover, the symbol detection duration considered at the Rx need to be in sync with the odor emission duration, i.e., the symbol time slot, as well to avoid errors resulting from decoding symbols over an incorrect symbol period. Furthermore, in order to minimise the latency in the OMC system, the receiver should be able to swiftly determine the received odor concentration over a symbol period. Instead of benzene, use of other odorants or odor mixtures for conveying the information could also be explored. Lastly, research is also required to investigate into other ICT aspects of an OMC system such as minimum SNR requirements, novel $M$-ary modulation schemes, error correction codes, multiple access techniques.
\bibliographystyle{IEEEtran}
\bibliography{IEEEabrv,main}
\end{document}